\documentclass[preprint]{revtex4}

\usepackage{graphicx}
\begin{document}
% You should use BibTeX and revtex.bst for references
\bibliographystyle{revtex}

% number  and your conference paper identification number on the
% title page in preprint mode. Multiple \preprint commands are allowed.
\preprint{IIT-HEP-01/7}

\title{An Asymmetric {\boldmath $e^+e^-$} Collider at the {\boldmath $\psi^{\prime\prime}$}\footnote{Submitted to {\sl Proc.\ Snowmass Summer Study on the Future of Particle Physics}, Snowmass, Colorado, 
June 30 -- July 21, 2001.}}
% Optional argument for running titles on pages
%\title[]{}



\author{Daniel M. Kaplan}
\email[]{kaplan@fnal.gov}
%\homepage[]{www.iit.edu/~kaplan}
%\thanks{}
\affiliation{Illinois Institute of Technology,
Chicago, IL 60616}

\author{Bruce C. Brown}
%\email[]{nelson@}
%\homepage[]{www.iit.edu/~kaplan}
%\thanks{}
\affiliation{Fermilab, Batavia, IL 60510}

\author{Harry N. Nelson}
%\email[]{nelson@}
%\homepage[]{www.iit.edu/~kaplan}
%\thanks{}
\affiliation{University of California at Santa Barbara, Santa Barbara, CA 93106}

%\collaboration{Muon Collaboration}

%\date{\today ~~D~R~A~F~T}

\begin{abstract}
A highly-asymmetric ``$\psi^{\prime\prime}$ factory" may be the best approach for studying $D^0{\overline D}{}^0$ mixing.

\end{abstract}
% \pacs{}

%\maketitle must follow title, authors, abstract and \pacs
\maketitle

%\section{Motivation}

The Standard Model predicts extremely small mixing between the $D^0$ and its antiparticle ${\overline D}{}^0$, thus $D^0{\overline D}{}^0$ mixing is potentially a window on new physics~\cite{Nelson}. Tantalizing hints from CLEO~\cite{CLEO} and FOCUS~\cite{FOCUS} that $D^0{\overline D}{}^0$ mixing may be on the verge of detectability in current experiments suggest that a dedicated experiment to study this phenomenon could be worthwhile. Photoproduction experiments are at the limit of statistics, and circular $e^+e^-$ colliders are systematically limited. While hadroproduction experiments such as BTeV could obtain orders of magnitude more reconstructed $D^0$ decays than either FOCUS or CLEO~\cite{BTeV}, they are likely to have poor efficiency at the short proper times where the mixing effect is largest. 

In principle $D^0$  mixing can be sought both in hadronic and in semileptonic $D^0$ decay modes~\cite{Morrison}. While the hadronic modes are better constrained (no missing neutrals) and have higher statistics, they have systematic uncertainty due to the difficulty of untangling mixing from doubly Cabibbo-suppressed decay, which leads to the same final states.
As at the $B$ factories, the decay $\psi^{\prime\prime}\to D^0{\overline D}{}^0$ has the appealing feature that the quantum numbers of the initial state forbid doubly Cabibbo-suppressed decays. This feature could be exploited at the proposed~\cite{CESR-c} CESR-$c$ facility, but with relatively low luminosity, since the $\psi^{\prime\prime}$ mass is lower than optimal for a ring the size of CESR. In a symmetric  $e^+e^-$ collider  set at $\sqrt{s}=m_{\psi^{\prime\prime}}$, there is also appreciable background from continuum events, which contributes systematic uncertainty.

%As suggested at Snowmass by H. Nelson, a 
A {\em highly}-asymmetric $e^+e^-$ $\psi^{\prime\prime}$ factory could be the solution to these problems. Consider, for the sake of discussion, collisions between a 50\,GeV positron beam (say, from the SLAC linac) and a high-intensity, low-energy electron beam. We require 
\begin{equation}
\sqrt{s}=m_{\psi^{\prime\prime}}=3770\,{\rm MeV}\approx\sqrt{2E_1E_2(1-\beta_1\beta_2\cos{\theta})}\,.
\label{eq:roots}
\end{equation}
With a crossing angle $\theta=90^\circ$ and $E_1=50\,$GeV, Eq.~\ref{eq:roots} is satisfied for $E_2=142\,$MeV. Such electron energy can be inexpensively produced by a small linac, however, achieving the required luminosity ${\cal L}\sim10^{33}$\,cm$^{-2}$s$^{-2}$ may require low-energy-beam intensity that is impractical for a conventional linac. The ``energy-recovery" linac may offer a practical solution~\cite{Ben-Zvi}. Another  possibility that has been considered is a ``proof-of-principle" laser-plasma-acceleration linac~\cite{Shvets}.

The aim in laying out such a facility would be kinematics for the decaying $D$  meson similar to those in a fixed-target experiment. The resulting high proper-decay-time precision and background suppression have been established repeatedly in experiments at Fermilab (Fig.~1). The large crossing angle assumed above should facilitate placement of vertex detectors close to the interaction point as in fixed-target experiments, albeit with a gap for passage of the high-energy beam, an arrangement that was used sucessfully in Fermilab E789~\cite{E789}. We hope to explore this idea further in the future.

\begin{figure}
%\vspace{3.5in}
%\hspace{-3in}
\scalebox{0.45}{\includegraphics{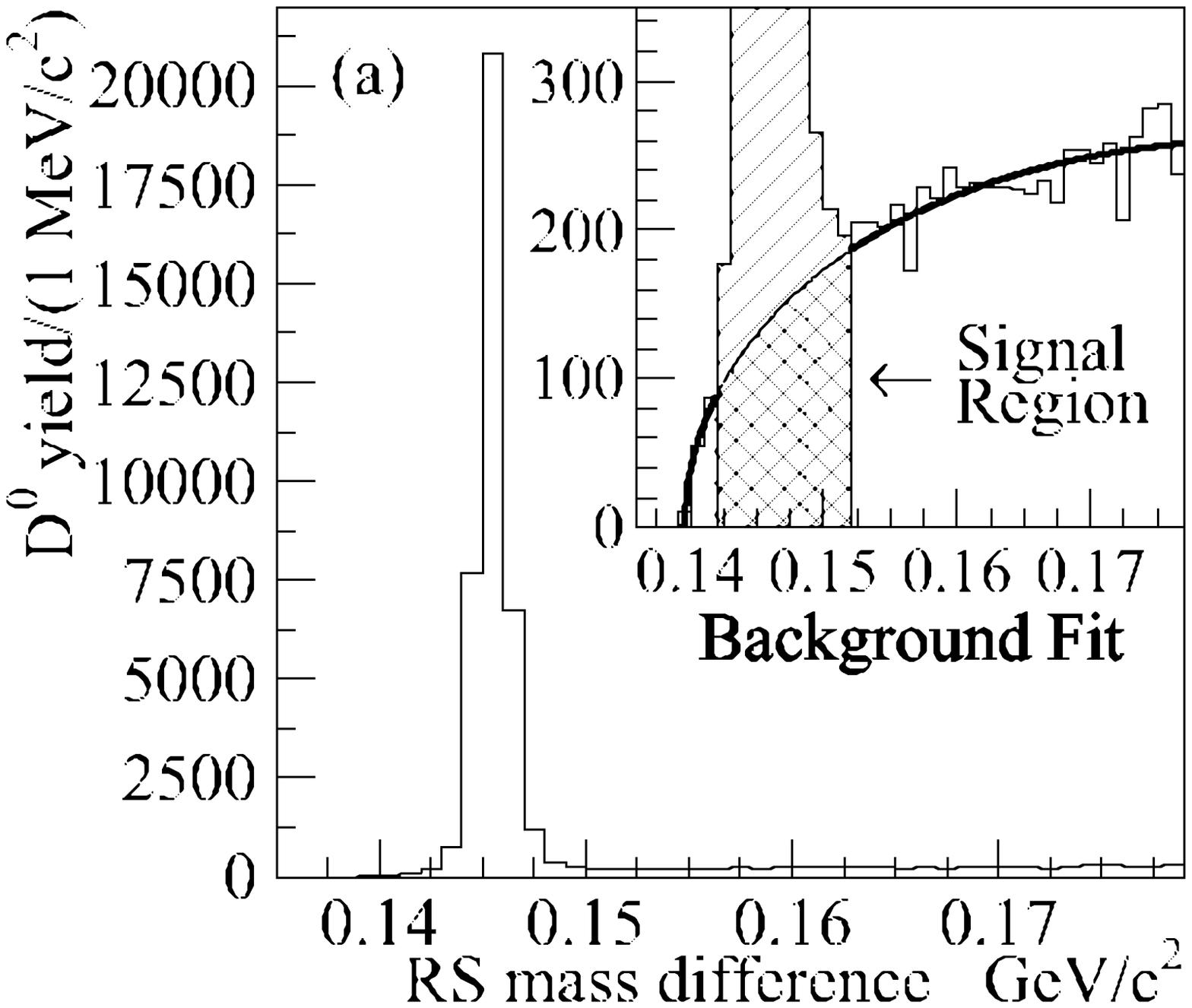}}
\hspace{.25in}
\scalebox{.5}{\includegraphics{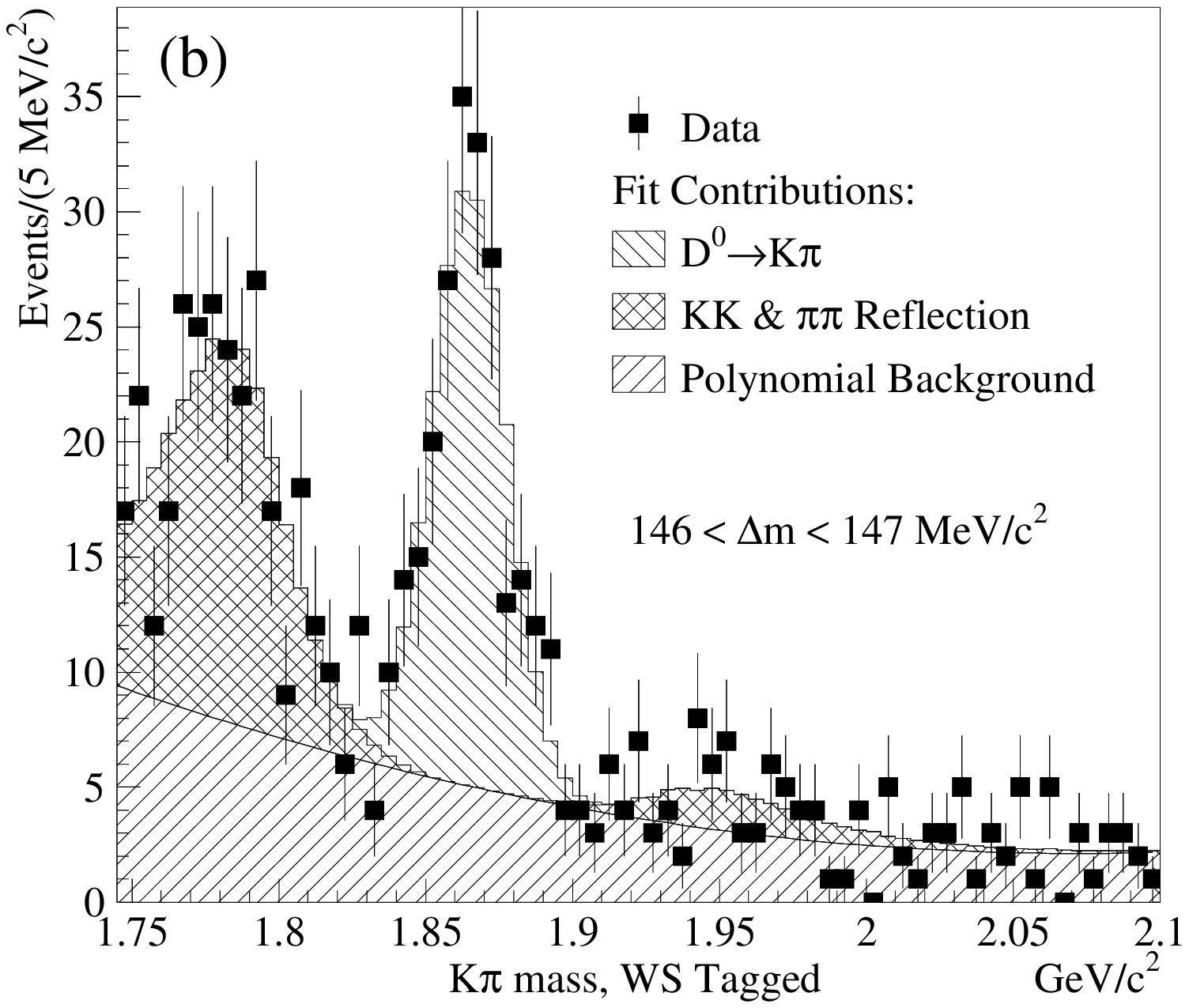}}
%\hspace{-6.7in}
\caption{Figures from \cite{FOCUS} showing cleanliness of FOCUS $D^0$ samples both for a) Cabibbo-allowed and b) doubly Cabibbo-suppressed decays.}
%\vspace{2.5in}

\end{figure}

\begin{acknowledgments}
The authors thank their various funding agencies, including the U.S. Dept.\ of Energy, the Illinois Dept.\ of Commerce and Community Affairs, and the Illinois Board of Higher Education, as well as a small black bear, the sighting of whom by the side of the road on the way down the hill early one morning led to a conversation between HNN and DMK without which this paper would not have been possible.
\end{acknowledgments}

\end{document}